# Manifestations of Spinodal Decomposition into Dilute Pd$_{1-x}$Fe$_x$ "Phases" in Iron-Implanted Palladium Films: FMR Study

A.I. Gumarov [1,2,a)], I.V. Yanilkin [1], A.A. Rodionov [1], B.F. Gabbasov[1], R.V. Yusupov [1], M.N. Aliyev [3], R.I. Khaibullin [2] & L.R. Tagirov [2,4]

[1] Institute of Physics, Kazan Federal University, 420008 Kazan, Russia
[2] Zavoisky Physical-Technical Institute of RAS, 420029 Kazan, Russia
[3] Baku State University, AZ-1148 Baku, Azerbaijan
[4] Tatarstan Academy of Sciences, Institute of Applied Research, 420111 Kazan, Russia

[a] Corresponding author: amir@gumarov.ru

## Abstract

Palladium-iron alloys produced by high-dose implantation of iron ions into epitaxial palladium films were investigated with the ferromagnetic resonance (FMR) and vibrating sample magnetometry (VSM) techniques. The samples reveal distinct multiple FMR responses depending on the dose of iron ion implantation. The post-implantation annealing at 770 K does not bring the implanted films to a homogeneous solid solution state, as might be expected from the Pd-Fe phase diagram. On the contrary, the system approaches a stable state composed of several magnetic phases. FMR spectra exhibit an angular behavior specific for a stack of interacting magnetic layers. This observation, correlated with the magnetometry data, indicates that the palladium-iron binary alloy has a previously unknown tendency towards spinodal decomposition into isostructural phases with well-defined iron concentrations and, accordingly, with different temperatures of ferromagnetic ordering and saturation magnetizations.

## 1. Introduction

Palladium-iron (Pd$_{1-x}$Fe$_x$) alloy is an unusual material where ferromagnetism sets in already at small, of the order of 1 at.%, amount of Fe atoms dissolved in palladium matrix [1,2]. The recent interest in Pd$_{1-x}$Fe$_x$ alloys originates from their potential application in superconducting spintronics, where palladium-rich compositions with $x < 0.10$ serve as a weak ferromagnetic link in magnetic Josephson junctions (MJJ) for ultrahigh-speed cryogenic devices [3-8]. The desire to achieve long superconducting coherence length (about 1-2 nm for elemental ferromagnets like Fe or Ni [9-11]) was a driving force for a search for low-temperature, weak ferromagnets like $Pd_{1-x}Fe_x$ (coherence length of 15-20 nm) to soften the deposition requirements to ferromagnetic (F) layers in MJJs. Simultaneously, it diminishes the destructive influence of the interface roughnesses on the coherence of supercurrent transport across the magnetic link. However, the non-stoichiometric solid solution like $Pd_{1-x}Fe_x$ alloy brings the issue of magnetic scattering [12] that introduces an additional attenuation of the superconducting pairing function and hence the reduction of the MJJ critical current. There is a controversial range of opinions on magnetic homogeneity of Pd-rich $Pd_{1-x}Fe_x$ alloys that includes clustering [13-17] and formation of $Pd_3Fe$ phase nanograins [18,19] in thin $Pd_{1-x}Fe_x$ films deposited with the magnetron sputtering. Magnetic homogeneity is claimed for the molecular-beam epitaxy (MBE) grown films [20-23] in the range of the iron contents of $x = 0.01$-$0.07$.

In our recent paper [24], we applied ion-beam implantation of iron into epitaxial palladium films to obtain dilute $Pd_{1-x}Fe_x$ alloy. The ion implantation technique is widely used for doping semiconductors and microchip fabrication [25,26]. The conventional procedure includes two main steps: irradiation with the dopant ions and subsequent annealing to heal radiation defects and restore crystallinity. In the course of the annealing, the initially inhomogeneous distribution of dopant [27] redistributes forming p-doped and n-doped regions of transistors. A similar strategy was employed in our sample preparation procedures, however, X-ray photoemission with depth profiling showed that the uniform iron-doping of palladium films was not achieved [24]. Moreover, according to our magnetometry and preliminary ferromagnetic resonance (FMR) data, a multiphase magnetic system with one to three components depending on the implantation dose was formed. This has not been expected from the binary phase diagram of Pd-Fe [28,29] where the homogeneous solid solution is expected on the palladium-rich side of the diagram. The hypothesis on spinodal decomposition during ion implantation was proposed in [24] based on the concepts outlined in [30–32]. The scenario of separation into large-scale, extended structures, when the energy difference of the precipitated phases is small [30], seems to fit the conditions of our experiment. In this paper, we report on the results of the extended FMR studies of the iron-implanted epitaxial palladium films to shed light on the structure of the stratified state. Our data provide experimental evidence in favor of decomposition into laminar structure along the iron concentration gradient directed perpendicular to the film surface.

## 2. Experimental section

### 2.1. Sample preparation and characterization

Epitaxial, high purity Pd (99.95%, *EVOCHEM GmbH*, Germany) thin films grown on epi-polished MgO (001) single-crystal substrate (*Crystal GmbH*, Germany) were used as the starting materials. The three-step procedure of the synthesis is described in detail in Ref. [20]. The $^{56}$Fe$^+$ ion implantation had been carried out with the *ILU-3* accelerator at a fixed ion energy of 40 keV and ion currents of 2-3 $\mu$A as measured at the sample. Three 10×5 mm$^2$ samples with the irradiation doses of 0.5×10$^{16}$, 1.0×10$^{16}$ and 3.0×10$^{16}$ ions/cm$^2$ were prepared varying the irradiation time. Each as-implanted sample was cut into pieces, and part of them was annealed in a vacuum (heating rate 15°C/min, then keeping at 770 K for 2 hours and cooling free to RT). We leave the same sample labels S0_5, S1_0, and S3_0 for as-implanted and S0_5a, S1_0a, and S3_0a for the annealed ones as were used in Ref. [24]. The doses yielded the samples with the mean iron contents of $\bar{c}_{Fe} \approx$ 2.5, 3.5, and 7.5 at.% in the implanted layer, respectively. Film thicknesses are given in Table 1. Further details on the characterization of the samples can be found in Ref. [24]. In the following, it is important to note that X-ray diffraction studies have shown that the Fe-implanted and subsequently annealed palladium films are single crystalline and epitaxial.

### 2.2 Ferromagnetic resonance measurements

For FMR measurements the specimens with a size of 2×2 mm$^2$ were cut from the samples. Here, we report on the properties of the annealed samples. FMR spectra were recorded (*Bruker ESP300 spectrometer*, X-band frequency $\nu \sim$ 9.5 GHz) in the in-plane and out-of-plane geometries (see insets to Fig. 1) in the temperature range of 20-300 K. During the in-plane measurements, the magnetic component of the microwaves was perpendicular to the film, whereas the external DC magnetic field was rotated in the film plane (azimuthal angle $\varphi_H$ was varied, see Fig. 1a). For the out-of-plane measurements, the magnetic field component of the microwave laid in the film plane, whereas the external DC magnetic field was rotated from the film plane (polar angle $\theta_H = 90°$) toward the film normal ($\theta_H = 0°$, see Fig. 1b).

The angular evolution of FMR spectra for the S3_0a sample (the dose is $3.0\times10^{16}$ ions/cm$^2$) is presented in Fig. 1. The spectra are the field-derivatives of the absorbed power and possess an asymmetric lineshape (the low- and high-field peaks have different amplitudes). To extract the resonance field and the linewidth with high accuracy, the spectra were fit with a model representing a mixture of the absorption and dispersion, $dP/dH \propto \sum_{i=1}^{3} d/dH[\chi_i''(H) + \alpha\chi_i'(H)]$, where the actual value of parameter $\alpha \in [0,1]$ depends on the electromagnetic screening conditions by the conducting film and/or a substrate [33]. For metallic films much thinner than the skin-depth, the value of $\alpha$ is much smaller than 1 (it is just our case). The Lorentzian shape was found to fit well the in-plane and the out-of-plane FMR spectra. At least three resonance lines can be identified on both the in-plane and out-of-plane records made at the temperature of 20 K for the S3_0a sample (see Fig. 1).

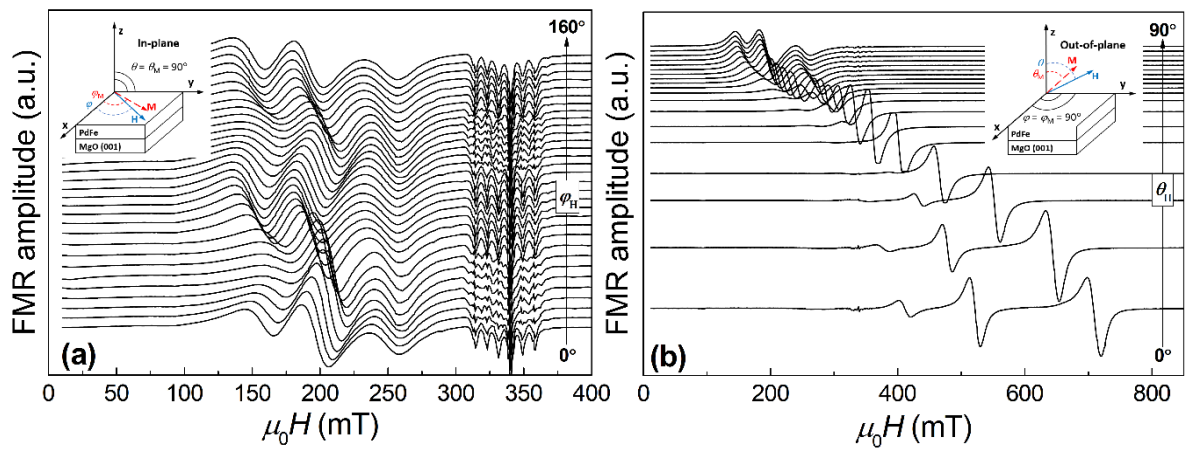

**Figure 1.** Angular evolutions of the FMR spectra of the S3_0a sample recorded in the in-plane (a) and out-of-plane (b) geometries of measurements; ν = 9.416 GHz, $T$ = 20 K.

Angular dependence of the resonance field in the out-of-plane geometry (Fig. 2a) demonstrates a conventional thin-film behavior [34,35]. Because of the demagnetization field, the resonance field in the normal direction to the film plane is shifted to high fields (up to about 7.4 kOe), whereas the resonance field in the tangential direction is shifted to low fields (below 3 kOe) relative to the electron spin resonance fields of the paramagnetic 3d-impurities. This resembles very much the angular behavior of the FMR field for resonance in easy-plane magnetic multilayers with interlayer exchange coupling [36].

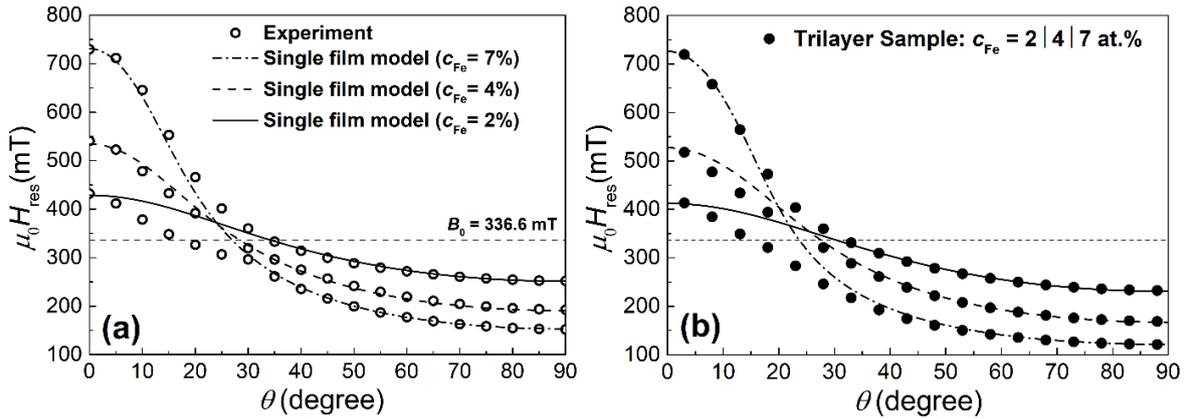

**Figure 2.** (a) Angular dependences of the FMR fields for resonance: (a) of the S3_0a sample and (b) of the MBE synthesized trilayer (see text); both –for the out-of-plane geometry of measurements. The FMR spectra were recorded at 20 K.

The evolution of the FMR spectra of the S3_0a sample with temperature is shown in Fig. 3a. Upon lowering the temperature, three FMR lines ($L1$, $L2$, and $L3$) emerge sequentially one after the other, each gradually shifting to the higher fields. This could be another indication of three distinct ferromagnetic components occurrence in the sample with different Curie temperatures and saturation magnetizations. Temperature dependence of the saturation magnetic moment for the S3_0a sample is shown in Fig. 3b. At a first glance, there are no obvious inflections or non-monotonicity of the curve that would indicate the presence of multiple magnetic phases in the sample.

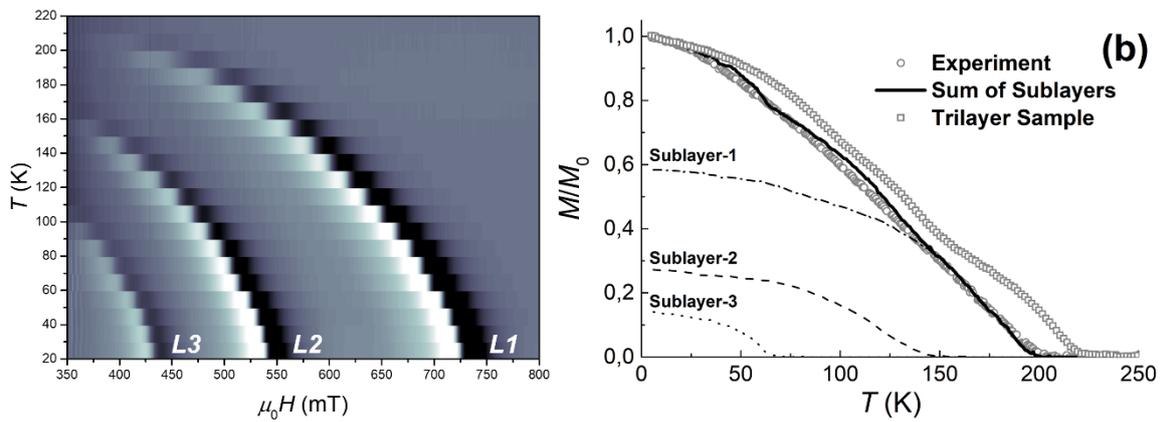

**Figure 3.** (a) Evolution of the FMR spectrum of the S3_0a sample with temperature. The spectra were recorded in the out-of-plane geometry with the magnetic field applied along with the film normal, $\nu = 9.416$ GHz. (b) Temperature dependence of the saturation magnetic moment for the S3_0a sample (open circles) and its representation by a sum of three homogeneously magnetized noninteracting layers (see text); analogous temperature dependence is shown for the MBE grown trilayer structure (open squares).

The S0_5a and S1_0a samples exhibit similar behavior, with the mid-dose S1_0a sample showing the transition from one to two FMR lines upon lowering the temperature, while the low-dose S0_5a sample showing a single FMR line in the entire temperature range where the line is detected. The FMR spectra of these samples are shown in Fig. 4 in comparison with that for the S3_0a sample. The resonance fields and linewidths of the FMR lines from Fig. 4 are collected in Table 1.

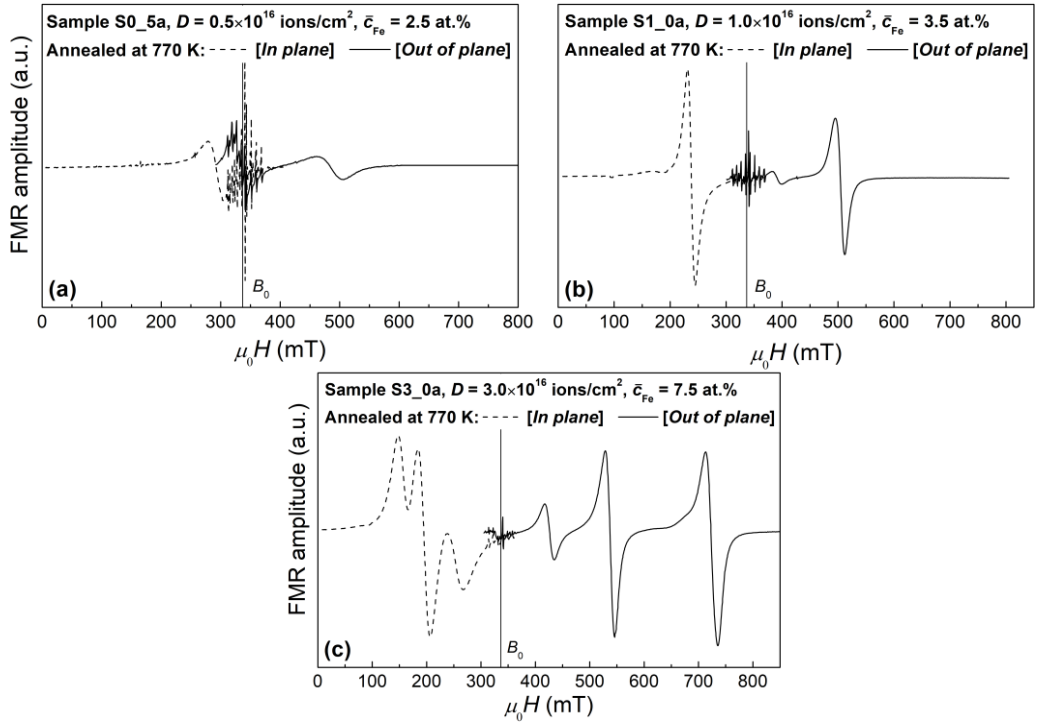

**Figure 4.** FMR spectra of the S0_5a, S1_0a, and S3_0a in the magnetic field directed along the normal and parallel to the plane of the sample. All FMR spectra were recorded at 20K.

**Table 1.** FMR linewidths for the geometry of the out-of-plane measurement at 20 K.

| Method | Sample label | Thickness (nm)* | Resonance field (mT) | | | FMR linewidth (mT) | | |
|---|---|---|---|---|---|---|---|---|
| Ion Implantation | S3_0a | 63 | 427 | 539 | 727 | 14.6 | 14.0 | 17.5 |
| | S1_0a | 50 | 392 | 505 | | 14.4 | 13.0 | |
| | S0_5a | 37 | 333 | 488 | | 31.7 | 34.4 | |
| Molecular-beam epitaxy | $Pd_{0.98}Fe_{0.02}$ | 20 | 397 | | | 11.3 | | |
| | $Pd_{0.96}Fe_{0.04}$ | 20 | | 512 | | | 7.8 | |
| | $Pd_{0.93}Fe_{0.07}$ | 20 | | | 726 | | | 7.8 |
| | $Pd_{0.98}Fe_{0.02}$ / $Pd_{0.96}Fe_{0.04}$ / $Pd_{0.93}Fe_{0.07}$ | 20/20/20 | 423 | 528 | 730 | 16.3 | 11.7 | 11.2 |

*The accuracy of the thickness measurements is ± 0.5 nm.

## 3. Discussion

Observation of multiple FMR lines indicates the formation of several magnetic phases in Fe-implanted palladium films. Indeed, for magnetically soft ferromagnetic films, the contribution of magnetocrystalline anisotropy to the resonance field can be neglected in the first approximation. Then, the resonance field in the direction of the normal to the thin film plane reads: $B_{res}^{\perp} = B_0 + 4\pi M_{eff} > B_0$, and in the direction parallel to the film plane $B_{res}^{\parallel} < B_0$ is a solution of the equation, $B_0^2 = B_{res}^{\parallel}(B_{res}^{\parallel} + 4\pi M_{eff})$, where $B_0 = 2\pi \nu/\gamma$, $\nu$ is the spectrometer operating frequency, $\gamma$ is a gyromagnetic ratio, $M_{eff}$ is the effective saturation magnetization of the film (see, for example, page 200 of Ref. [36]). Figures 2 and 4 demonstrate that the angular behavior qualitatively reproduces the FMR response of thin ferromagnetic films with the easy-plane shape anisotropy [21,23,34-36], and according to the above expressions, every FMR line in the spectrum can be assigned to a material fraction with particular $M_{eff}$ and 2D-morphology (the deviations from single-layer traces in Fig. 2 are most probably caused by the interaction between the "layers"). This, in turn, could be a manifestation of the Pd-Fe system separation into certain compositions as its magnetization correlates with local iron content. Three FMR lines were observed already in the as-implanted S3_0 sample [24] meaning that the separation occurred in the course of the implantation, while the annealing just brought the system to the achievable equilibrium leaving the number of FMR signals intact.

In our opinion, this can take place due to a kind of spinodal decomposition [30–32] occurring under non-equilibrium conditions in the course of the ion implantation [24]. Probably, the inhomogeneous profile of the iron distribution in the implanted palladium films leads to a formation of a laminar multiphase magnetic system with different temperatures of ferromagnetic ordering and saturation magnetizations for each phase (sub-layer). This explains why the FMR lines emerge one after the other upon lowering the temperature (see Fig. 3a). At the same time, different temperatures of ferromagnetic ordering also directly correspond to definite compositions of the Pd$_{1-x}$Fe$_x$ alloy [23].

Importantly, the resonance fields for the FMR lines of the samples S3_0a, S1_0a, and S0_5a implanted to different doses of iron are near identical though the patterns are different, see Fig. 5. This finding supports a formulation of a simple model of a spinodal decomposition into the isostructural laminar system [30], each sub-layer possessing a certain, quite well-defined concentration of iron in palladium (see a sketch in Fig. 6a based on the depth profiling of the iron concentration across the S3_0a sample thickness in Ref. [24]). The system evolves from a non-equilibrium state created by ion implantation to its energetically favored state by the formation of several spatially-extended (macroscopic) "stable phases" [30].

Once the correspondence between the saturation magnetic moment and the iron concentration (*i.e.* the $M_s(x)$ dependence) is established for the MBE grown $Pd_{1-x}Fe_x$ alloy films [23], we can estimate the iron content in these $Pd_{1-x}Fe_x$ "stable phases" in the implanted and subsequently annealed samples as $x \approx 0.025$, $x \approx 0.042$, and $x \approx 0.069$. It is of course an approximate estimate because $M_{eff}(x)$ extracted from the FMR measurements is an averaged value over a sub-layer, while $M_s(x)$ is a property of the MBE grown films having maximum homogeneity; the possible exchange coupling between the sub-layers in S1_0a and S3_0a samples have not been taken into account. Moreover, there is some scatter of resonance field values for the groups of lines, highlighted in Fig. 5, see Table 1. Additionally, we have estimated the iron concentration in each stable phase from the temperatures corresponding to the appearance of the FMR signals (Fig. 3a). Results of these estimations are shown in Table 2.

**Table 2.** Estimates of the iron concentrations in the ferromagnetic phases in implanted palladium films from FMR and dc magnetization data.

| Sample label | Sublayer | Iron concentration (at.%) defined from: | | |
|---|---|---|---|---|
| | | $H_{res}$ (at 20 K) | $H_{res}$ (T) | $M$ (T) |
| S3_0a | 1 | 2.5 | 2.5 | 1.6 |
| | 2 | 4.2 | 5 | 4.7 |
| | 3 | 6.9 | 7 | 7 |
| S1_0a | 1 | 2 | 1.2 | 2.3 |
| | 2 | 3.7 | 3.4 | 3.8 |
| S0_5a | 1 | 3.4 | 1.8 | 2.1 |

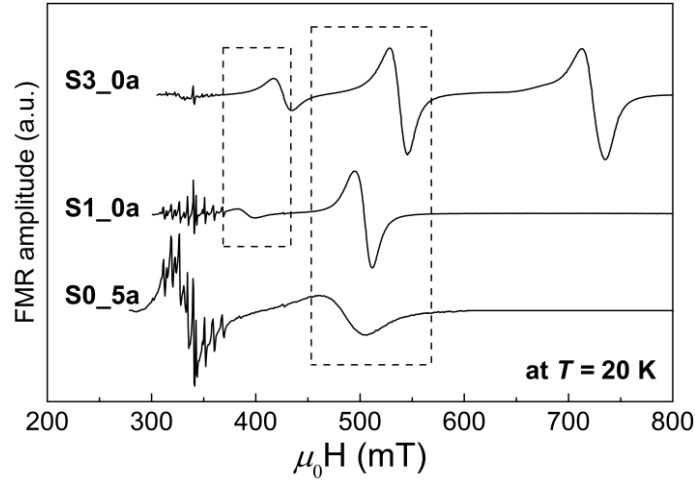

**Figure 5.** FMR spectra for all three Fe-implanted palladium films after the annealing for comparison of the resonance fields.

It is natural to think that these sub-layers with the iron contents estimated above contribute to the integral $M(T)$ presented in Fig. 3b. Indeed, the net magnetization temperature dependence is expressed as $M(T) = aM_1(T) + bM_2(T) + cM_3(T)$, where $M_i(T)$ is a magnetic moment of the *i*-th sub-layer, taken from Ref. [23] for the MBE-grown $Pd_{1-x}Fe_x$ films with $x = 0.016$, $x = 0.047$, and $x = 0.07$, and presented in Fig. 3b by the black dashed line, reproduces experimental $M(T)$ quite well. Here, only *a*, *b,* and *c* coefficients were adjusted for better correspondence. The above analysis gives similar results also for the S1_0a sample showing the presence of two ferromagnetic phases. The iron concentrations obtained from these deconvolutions for all three implanted samples are collected in Table 2.

To verify the spinodal decomposition hypothesis, we prepared an artificial layered sample that semiquantitatively mimics the tentative model of sample S3_0 depicted in the inset to Fig. 6a. Starting from epi-polished single-crystal MgO (001) substrate, three layers of $Pd_{0.98}Fe_{0.02}$, $Pd_{0.96}Fe_{0.04}$, $Pd_{0.93}Fe_{0.07}$ were grown epitaxially one by one (see details of the growth and characterization in Ref. [20]) resulting in fully epitaxial MgO/$Pd_{0.98}Fe_{0.02}$/$Pd_{0.96}Fe_{0.04}$/$Pd_{0.93}Fe_{0.07}$ stack. For simplicity, all layers were grown ~ 20±1 nm thick, so the total thickness of the stack was equal to ~ 60 nm near-identical to that of the S3_0a sample.

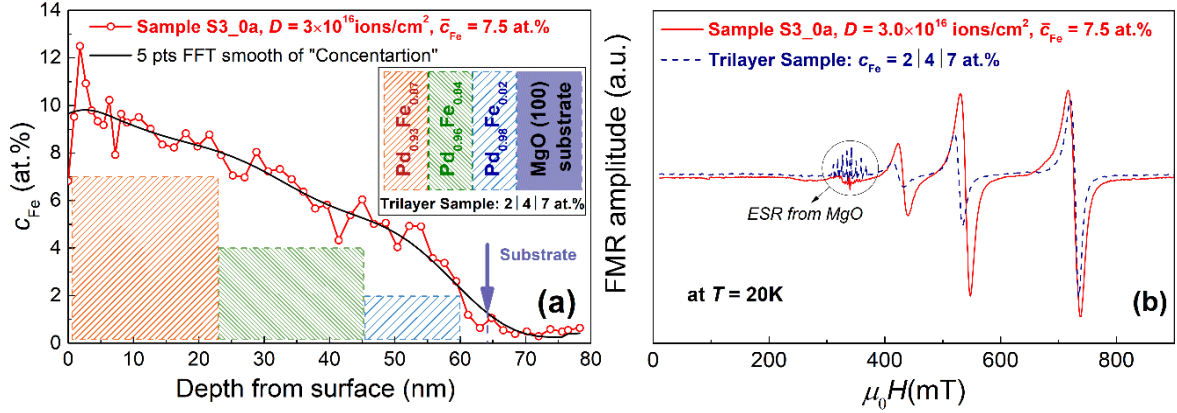

**Figure 6.** (a) A qualitative model of sample S3_0a of iron-implanted palladium film decomposition to sub-layers with different iron contents; the solid black line is obtained by smoothening of the concentration experimental points using FFT algorithm. Inset: a sketch of the MBE-grown trilayer; (b) FMR spectra of the annealed S3_0a sample (solid line) and MBE grown trilayer sample (dash line)

FMR spectra of the annealed S3_0a sample and the MBE grown trilayer sample for the field direction normal to the sample plane are presented in Fig. 6b. They reveal great similarity in the field for resonance and some discrepancy in the amplitude of signals. The latter, however, could be corrected by redistributing the layer thicknesses in favor of $Pd_{0.96}Fe_{0.04}$ and $Pd_{0.98}Fe_{0.02}$ layers while keeping the total thickness fixed. The angular dependence of the FMR field for resonance for the MBE trilayer, Fig. 2b, is also very close to that for the S3_0a sample, Fig. 2a. In our opinion, the observed close similarity in FMR and magnetization data for the iron-implanted and synthesized trilayer samples support the hypothesis of the spinodal decomposition in iron-implanted palladium epitaxial films during ion implantation and subsequent thermal annealing.

## 4. Conclusion

To summarize, extensive FMR and magnetometry studies of $Pd_{1-x}Fe_x$ alloy films obtained by high-dose Fe-ion implantation into epitaxial Pd-films have been performed. Multiple resonances are observed in the FMR spectra, each emerging at a particular temperature. This feature develops in the course of the implantation and is resistant to annealing at high temperatures.

Spinodal decomposition of iron-implanted film into a laminar magnetic structure in an otherwise crystallographically homogeneous film is proposed to be at the origin of the multi-line magnetic response. The angular dependence of the FMR field for resonance complies with this simple model with each sub-layer possessing certain, well-defined $Pd_{1-x}Fe_x$ composition.

The hypothesis was verified using the artificial structure comprising three $Pd_{1-x}Fe_x$ layers with different iron contents. The FMR spectra and magnetometry data are qualitatively identical to that of the iron-implanted Pd films with only a minor quantitative discrepancy taking place. All the collected data are consistent with the hypothesis of spinodal decomposition of the intrinsically non-equilibrium distribution of the delivered dopant to isostructural, laterally extended laminar magnetic sub-structures – the property which is absent in the conventional phase diagram of the binary Pd-Fe alloy.

From the applications point of view, this previously unknown tendency of the $Pd_{1-x}Fe_x$ binary system in the palladium-rich ($x < 0.10$) composition determines the ranges of intrinsically homogeneous compositions suitable for the use in superconducting spintronic heterostructures.

**Declaration of Competing Interest**

The authors declare that they have no known competing financial interests or personal relationships that could have appeared to influence the work reported in this paper.

**Acknowledgments**

The authors are grateful to Golovchanskiy Igor (MISIS, Moscow) for a fruitful discussion of the results. Synthesis and analysis of the films were carried out at the PCR Federal Center of Shared Facilities of KFU with the running costs covered by the Program of Competitive Growth of Kazan Federal University.

**Funding**

This work was supported by the RFBR Grant No. 20-02-00981.

**Data available on request from the authors**

The data that support the findings of this study are available from the corresponding author upon a reasonable request.